%% file: kps.tex
\pdfoutput=1
\documentclass[useAMS,usenatbib]{mnras}

\usepackage[british]{babel}             
\let\la=\lesssim
\usepackage[T1]{fontenc}                
 \usepackage{graphicx}                   

\input{journal}

\usepackage{graphicx}
\usepackage{float}
\usepackage{upgreek}
\usepackage{amsmath}
\usepackage{hyperref}
\def\fdeg{\hbox{$.\!\!^\circ$}} 

\providecommand{\adsurl}[1]{\href{#1}{ADS}}
 
\newcommand{\angstrom}{\textup{\AA}}
\usepackage[T1]{fontenc}
\usepackage{aecompl}

\title[First results of the Kourovka Planet Search]{First results of the Kourovka Planet Search: discovery of transiting exoplanet candidates in the first three target fields}
\author[Artem Y. Burdanov et al.]{Artem Y. Burdanov$^{1}$\thanks{E-mail: burdanov.art@gmail.com}\thanks{Present address: Institut d'Astrophysique et G\'  {e}ophysique, Universit\'{ e} de Li\` {e}ge, all\' {e}e du 6 Ao\^ {u}t 17, 4000 Li\` {e}ge, Belgium}, Paul Benni$^{2}$, Vadim V. Krushinsky$^{1}$, Alexander A. Popov$^{1}$, 
\newauthor 
Evgenii N. Sokov$^{3,5}$,
Iraida A. Sokova$^{3}$,
Sergei A. Rusov$^{3}$, 
Artem Yu. Lyashenko$^3$,
\newauthor
Kirill I. Ivanov$^{4}$,
Alexei V. Moiseev$^{5}$,
Denis A. Rastegaev$^{5}$, 
Vladimir V. Dyachenko$^{5}$,
\newauthor 
Yuri Yu. Balega$^{5}$,
\"Ozg\"ur Ba\c{s}t\"urk$^{6}$,
Ibrahim  \"Ozavc{\i}$^{6}$,
Damian Puchalski$^{7}$,
\newauthor
Alessandro Marchini$^{8}$,
Ramon Naves$^{9}$,
Stan Shadick$^{10}$
and Marc Bretton$^{11}$
\\
$^{1}$Ural Federal University, ul. Mira d. 19, Yekaterinburg, Russia, 620002\\
$^{2}$Acton Sky Portal (Private Observatory), Acton, MA, USA\\
$^{3}$Central Astronomical Observatory at Pulkovo of Russian Academy of Sciences, Pulkovskoje shosse d. 65, St. Petersburg, Russia, 196140\\
$^{4}$Irkutsk State University, ul. Karla Marxa d. 1, Irkutsk, Russia, 664003\\
$^{5}$Special Astrophysical Observatory, Russian Academy of Sciences, Nizhnij Arkhyz, Russia, 369167\\
$^{5}$Ankara University, Faculty of Science, Department of Astronomy and Space Science, TR-06100 Tandogan, Ankara, Turkey\\
$^{7}$Centre for Astronomy, Nicolaus Copernicus University, Grudziadzka 5, 87-100 Torun, Poland\\
$^{8}$Astronomical Observatory - DSFTA, University of Siena, Via Roma 56, 53100 Siena, Italy\\
$^{9}$Observatori Montcabrer, C/Jaume Balmes, 24, Cabrils, Spain\\
$^{10}$Physics and Engineering Physics Department, University of Saskatchewan, Saskatoon, SK, Canada, S7N 5E2\\
$^{11}$Baronnies Proven\c{c}ales Observatory, Hautes Alpes - Parc Naturel R\' egional des Baronnies Proven\c {c}ales, 05150 Moydans, France\\
}
\begin{document}


\pagerange{\pageref{firstpage}--\pageref{lastpage}} \pubyear{2016}

\maketitle

\label{firstpage}

\begin{abstract}
We present the first results of our search for transiting exoplanet candidates as part of the Kourovka Planet Search (KPS) project. The primary objective of the project is to search for new hot Jupiters which transit their host stars, mainly in the Galactic plane, in the \textit{$R_c$} magnitude range of 11 to 14~mag. Our observations were performed with the telescope of the MASTER robotic network, installed at the Kourovka astronomical observatory of the Ural Federal University (Russia), and the Rowe-Ackermann Schmidt Astrograph, installed at the private Acton Sky Portal Observatory (USA). As test observations, we observed three celestial fields of size $2\times2$~deg$^2$ during the period from 2012 to 2015. As a result, we discovered four transiting exoplanet candidates among the 39000 stars of the input catalogue. In this paper, we provide the description of the project and analyse additional photometric, spectral, and speckle interferometric observations of the discovered transiting exoplanet candidates. Three of the four transiting exoplanet candidates are most likely astrophysical false positives, while the nature of the fourth (most promising) candidate remains to be ascertained. Also, we propose an alternative observing strategy that could increase the project's exoplanet haul.  

\end{abstract}

\begin{keywords}
stars: planetary systems -- techniques: photometric
\end{keywords}

\section{Introduction}

As of the end of 2015, the existence of more than 1500 extrasolar planets in the vicinity of stars of different types has been reliably confirmed \citep{2011A&A...532A..79S, 2011PASP..123..412W}. Most of the newly found exoplanets were discovered through photometric observations of their transits or by analysing the radial velocity variations of the host stars (for a review of the history of the search for exoplanets and the main techniques of the discovery, see \cite{Perryman2012} and \cite{WrightGaudi2013}). Furthermore, most of the 4000 exoplanet candidates from the \textit{Kepler} mission are, in all likelihood, planets, too \citep{2015ApJS..217...31M}.

The anticipated number of exoplanets in our Galaxy is significantly larger than the number of the already discovered extrasolar planets, which renders their primary distribution similar to the 'tip of the iceberg'. In addition, the available distribution of exoplanets is distorted by the selection effects of each of the search methods. Therefore, each newly discovered exoplanet with reliably determined characteristics will increase the population of the known exoplanets and contribute to the further development of the planet population synthesis, this being the method by which the synthetic populations of exoplanets are compared with the observed ones \citep{2015IJAsB..14..201M}. The advancement of this method will shed light on some of the aspects related to the formation and evolution of extrasolar planets. On the other hand, the search for new transiting exoplanets is also necessary for determining a more accurate boundary between late M dwarfs, brown dwarfs, and gaseous giant exoplanets due to the similar size of these objects. 

\begin{figure}
\centering
\includegraphics[width=\columnwidth]{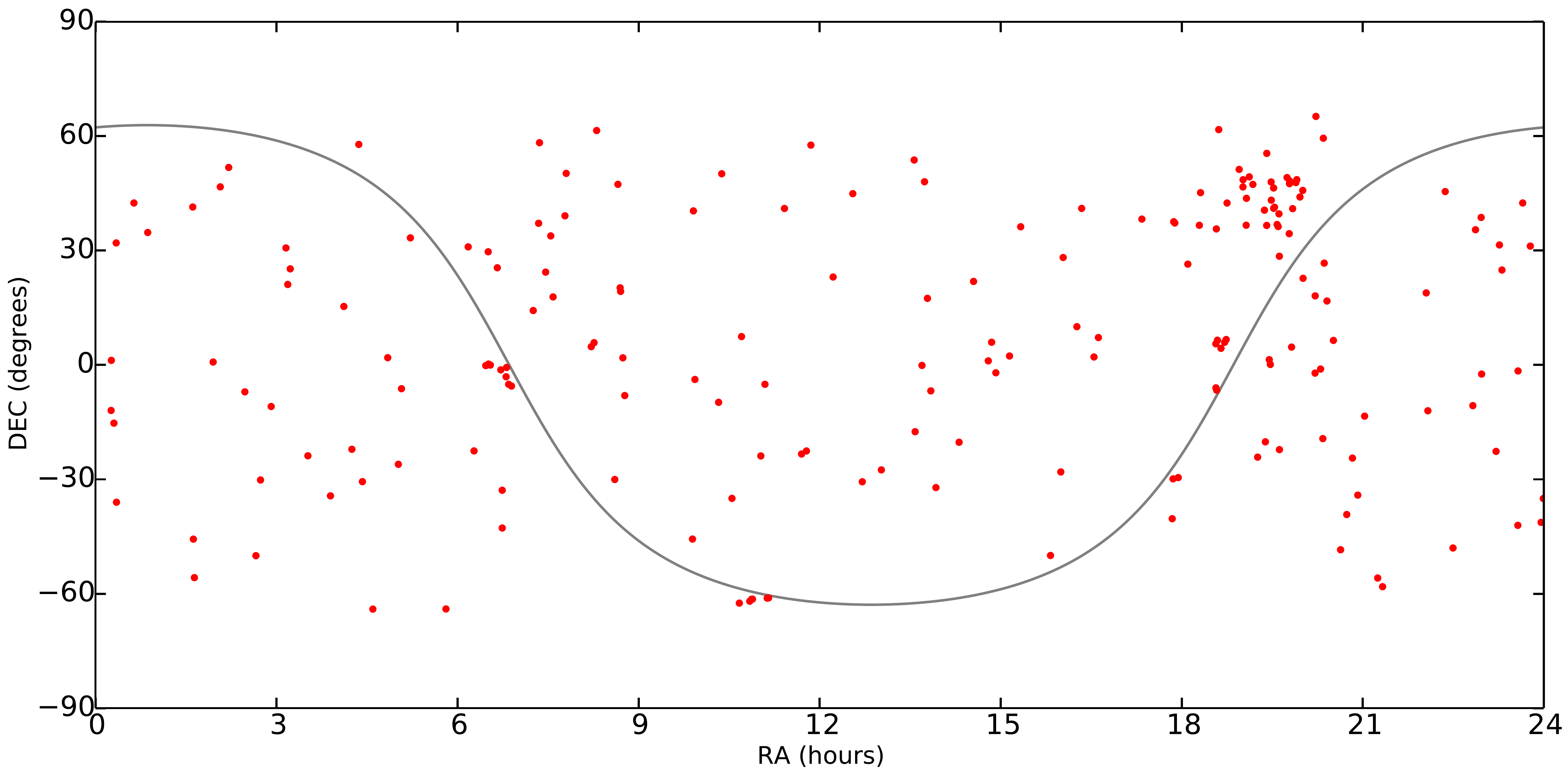}
\caption{Spatial distribution of the known transiting hot Jupiters, i.e. exoplanets with orbital periods less than 10 days and with the minimum mass greater than $0.5 M_J$ \citep{2011PASP..123..412W}. The Galactic plane is shown by the grey solid curve.}
\label{fig00}
\end{figure}

Ground-based photometric surveys, such as HATNet \citep{Bakos2004}, HATSouth \citep{2013PASP..125..154B}, KELT \citep{Pepper2007}, OGLE \citep{Udalski2003}, QES \citep{Alsubai2013}, TrES \citep{Alonso2004}, WASP \citep{2006PASP..118.1407P}, and XO \citep{McCullough2005}, along with space projects such as \textit{Kepler} \citep{Borucki2010} and \textit{CoRoT} \citep{Auvergne2009}, have discovered more than half of the known exoplanets through observations of a significant portion of the celestial sphere. However, the sky has not yet been covered in its entirety by such observations (see Fig.~\ref{fig00}). Most ground-based photometric surveys that focus on the search for transiting hot Jupiters are incapable of reliably recording transits for host stars fainter than 13 mag in the \textit{$R_c$} band and, most importantly, avoid observing dense fields in the Galactic plane. Thus, a large number of hot Jupiters transiting relatively bright stars are waiting to be discovered there. 

The Kourovka Planet Search (KPS) project was organised at the Kourovka astronomical observatory of the Ural Federal University (Yekaterinburg, Russia). One of the observatory's photo-electric telescopes with a two-star photometer had already participated in one of the first attempts to find exoplanets within the framework of the 1994--1999 TEP project (see \cite{2000ApJ...535..338D} and \cite{2008A&A...480..563D}). Our observations were made with the robotic MASTER-II-Ural telescope. This instrument forms part of the network of MASTER telescopes \citep{Lipunov2010}. The primary mission of the MASTER project is to carry out synoptic surveys of the sky, searching for various optical transients (supernovae, optical afterglow of gamma-ray bursts, and asteroids). A synoptic sky survey does not require high precision photometry, but the technical characteristics of the system make it possible to perform photometric observations of 11 to 14 mag stars with 1~percent precision in the \textit{$R_c$} band. For that reason, we used the MASTER-II-Ural telescope to study open clusters \citep{Popov2013} as well as to observe the transits of other known exoplanets \citep{Gorbovskoy2013,2015MNRAS.450.3101B}.

When testing the technique of high precision photometry with the MASTER-II-Ural  telescope, we decided to carry out a pilot photometric survey in search for transiting exoplanets. In 2012, we observed with the MASTER-II-Ural telescope a  region of the sky in the constellation of Cygnus. During 2013--2014, we carried out a survey of another region of the celestial sphere in the Cassiopeia constellation, which was also observed with the Celestron Rowe-Ackermann Schmidt Astrograph (RASA) telescope installed at the private Acton Sky Portal Observatory (Massachusetts, USA). A third region of the celestial sphere, in the Ursa Major constellation, was observed with the RASA telescope only.

The MASTER and RASA telescopes provide the optimal parameters for searching for such exoplanets transiting their host stars, with magnitudes ranging from 11 to 14 mag. The expected transit depth in this case is about 0.01--0.02 mag, and can be detected with 0.001 mag precision. The 1.8~arcsec~pixel$^{-1}$ image scale allows one to observe with high photometric precision the celestial regions of the Galactic plane with large concentrations of stars. Current ground-based wide-field surveys usually stay clear of dense portions of the Galactic plane to avoid the blending of stars and associated problems of photometric processing. The KPS survey is similar in image scale to the ongoing exoplanet photometric surveys BEST II \citep{2014PASP..126..227F, 2009A&A...506..569K} and ASTEP 400 \citep{2006ESASP1306..513F, 2010SPIE.7733E..4TD, 2010A&A...511A..36C}.  
 
This paper is organised as follows: Section \ref{INSTRUMENTS AND OBSERVATIONS} is devoted to the description of the instruments used and observations made; Section \ref{DATA REDUCTION} presents the observed data processing technique; Section \ref{CANDIDATES} deals with the detected transiting exoplanet candidates and with their additional photometric, spectroscopic, and speckle interferometric observations; in \hyperref[CONCL]{conclusion} we discuss the main results and the future of the project.  

\section{INSTRUMENTS AND OBSERVATIONS}\label{INSTRUMENTS AND OBSERVATIONS}

Observations of the first field of the KPS survey were made with the MASTER-II-Ural telescope only, installed at the Kourovka astronomical observatory ($\varphi=57\degr$~N, $\lambda=59\degr$~E). The telescope consists of a pair of Hamilton catadioptric tubes (400~mm, f/2.5) installed on one equatorial Astelco NTM-500 mount. Each tube is equipped with a $4098\times4098$ pixel Apogee Alta~U16M CCD yielding an image scale of 1.8~arcsec~pixel$^{-1}$ in a $2\times2$~deg$^2$ field. The average full width at half maximum (FWHM) of the stellar profiles is 3--4~arcsec. Each tube of the telescope has four filters: \textit{V, $I_c$}, P, and the 'red continuum' on the east tube and \textit{$R_c$, B}, P, and H~$\upalpha$ on the west tube. Filters \textit{BV$R_cI_c$} constitute a Johnson-Cousins system, and the P filters are polarised with perpendicular orientation. The 'red continuum' is a narrowband filter 5~nm wide transmitting radiation in the 645~nm wavelength region. The H~$\upalpha$ filter also has a width of 5~nm and is centred on the wavelength 656~nm. 

\par
\begin{table*}
\small
\caption{Parameters of observations made with the MASTER-II-Ural and RASA telescopes}
\label{tab:sets}
\centering
\begin{tabular}{ccccccc}\hline
Field & Coordinates & Telescope & Year of observation & Number of images & Exposures & Filter\\
& of centre (J2000) &\phantom{text} & \phantom{text} & \phantom{text} & s\\
\hline
TF1 & 20:30:00 +50:30:00 & MASTER-II-Ural & 2012 & 3600 & 50 & $R_c$\\
TF1 & 20:30:00 +50:30:00 & MASTER-II-Ural & 2012 & 1000 & 120 & $BV$\\
TF2 &  02:47:00 +63:00:00 & MASTER-II-Ural & 2013--2014 & 4400 & 50 & $VR_c$\\
TF2 & 02:47:00 +63:00:00 & MASTER-II-Ural & 2014 & 500 & 120 & $BI_c$\\
sub-TF2 & 02:47:00 +63:00:00 & RASA & 2014 & 8000 & 50 & $R_c$\\ 
sub-TF2 & 02:47:00 +63:00:00 & RASA & 2014 & 485 & 50 & $V$\\
TF3 & 11:00:00 +65:48:00& RASA & 2015 & 7000 & 50 & $R_c$\\
TF3 & 11:00:00 +65:48:00 & MASTER-II-Ural & 2015 & 50 & 120 & $BVR_cI_c$\\\hline
\end{tabular}
\end{table*}
\par

For our pilot observations we selected a Milky Way field in the Cygnus constellation with a high density of stars which does not overlap with the \textit{Kepler} field of view. The field was named TF1 (Target Field~\#1).The coordinates of the TF1 centre are $\upalpha_{\text{J2000}}=$ 20:30:00 $\updelta_{\text{J2000}}=$ +50:30:00. The choice of TF1 was based on two main factors: favourable observing conditions during the summer at the Kourovka observatory, and the absence of known exoplanets there.

The main TF1 observation set was completed during the short and bright summer nights from May to August~2012. We obtained 3600 frames with 50~s exposure times, but only in the $R_c$ band, because only one of the CCD cameras of the telescope was operable at the time. The time interval between consequent frames was about 1.5~min.  

After the initial data reduction, we performed follow-up observations of TF1: we acquired $R_c$-band data with 50~s exposure times, and \textit{B}- and \textit{V}-band frames with 120~s exposure times, to determine the colour indices of the stars from the input catalogue (description of the catalogue is provided in Section \ref{ftis2lc}). All in all, TF1 was observed for 90~h in the $R_c$ band (36 nights) with a 2.5~h average duration of an observing night.

The second target field (TF2) is located in a less dense zone of the Milky Way in the Cassiopeia constellation. The coordinates of the centre are $\upalpha_{\text{J2000}}=$ 02:47:00 $\updelta_{\text{J2000}}=$ +63:00:00. The choice of TF2 was based on the same principles as TF1.

The TF2 field was observed with the MASTER-II-Ural telescope for 100~h (43 nights) in 2013 and 2014. The major portion of the observed data consists of 4400 frames obtained simultaneously with the two tubes of the telescope in the \textit{V} and $R_c$ bands with 50~s exposure times. Likewise, we conducted \textit{B}- and $I_c$-band observations to determine the colour indices of the stars from our input catalogue. As in the case of TF1, the average observing night lasted for only 2.5~h due to the weather conditions at the telescope site.

All observations with the MASTER-II-Ural telescope were made in automatic mode. Prior to each observation night, dark frames were obtained with the necessary CCD temperature; under good weather conditions, every morning the telescope obtained twilight sky flat-field frames in the appropriate filter. 

In September 2014, a second telescope was involved in the KPS project. The Rowe-Ackermann Schmidt Astrograph (Celestron Inc., Torrance, CA, USA) is operated at the private observatory Acton Sky Portal, Massachusetts, USA ($\varphi=43\degr$~N, $\lambda=71\degr$~W). The telescope consists of a 279~mm diameter mirror with a focal length of~620 mm. Initially, a Celestron CGEM mount was used; it was later replaced by a Losmandy Titan mount (Hollywood General Machining, Los Angeles, CA, USA) to provide new observing regimes. The installed CCD camera is a $3352\times2532$ pixel SBIG ST-8300M (SBIG, Santa Barbara, CA, USA) which provides an image scale of 1.8~arcsec~pixel$^{-1}$ in a $1.2\times1.6$~deg$^2$ field. This FOV is almost twice smaller than the MASTER-II-Ural telescope FOV. Typical PSF FWHM values for stars are 2--3~arcsec. Despite the smaller FOV, this telescope benefits from clearer observing nights than those at the Kourovka observatory site.

The RASA telescope obtained 8000 frames containing the inner subregion of the TF2 field. The observations were carried out from early September 2014 to late November 2014, in the $R_c$ filter, with 50~s exposure times. The time interval between the frames was 1~min. Also, \text{V} band frames were obtained on one night. All in all, the TF2 field was observed with the RASA telescope in the R filter for 130~h (18 nights), averaging 7.2~h per night. 

The third field, TF3, is located in the Ursa Major constellation and contains the smallest number of stars of all the observed fields. This field is not part of the Milky Way, but was nonetheless observed as part of the test observation with the RASA telescope.  TF3 was observed with the RASA telescope only in the $R_c$ band, from January to April 2015. The TF3 field was observed for 115~h (21 nights), averaging 5.5~h per night. The MASTER-II-Ural telescope obtained  \textit{BV$R_cI_c$} band frames of TF3 for determining the colour indices of the stars.

The information about the obtained observation sets is summarised in table \ref{tab:sets}.

The volume of the obtained data amounted to several terabytes, which is not very large by modern astronomical standards, yet it required automated flow processing. The description of the data processing and analysis procedures is provided in the next section.

\section{DATA REDUCTION and SEARCH FOR VARIABLE STARS AND EXOPLANET TRANSITS}\label{DATA REDUCTION}

All obtained FITS files were transferred to the IT cluster of the Ural Federal University where they were processed using a computer with a Debian operating system.      

Prior to data processing, we had to filter some of the FITS files. Not all observations were made in optimal weather conditions due to operational instability of the Boltwood Cloud Sensor~II at the Kourovka observatory. To this end, we analysed the standard deviation of the pixel counts $\sigma_{\text{pix}}$ for each image. The value of $\sigma_{\text{pix}}$ varies smoothly from image to image on the photometric nights. In the presence of clouds, $\sigma_{\text{pix}}$ varies significantly, making it possible to discard the images obtained in poor weather conditions.

\subsection{From FITS Files to Light Curves}\label{ftis2lc}
 
For automated processing of the FITS files, we developed a {\scshape k-pipe} script using the Bash programming language. The script combines the procedures of photometric calibrations of frames in the {\scshape iraf} package \citep{Tody1986}, creation of the WCS header in the {\scshape astrometry.net} application~\citep{Lang2010}, aperture photometry in the {\scshape iraf} package, and differential photometry in the {\scshape astrokit} application~\citep{2014AstBu..69..368B}.

Initially, the {\scshape iraf} package was used for the photometric reduction of each frame, i.e. to subtract the master dark frame and divide by the master flat field frame. 

The physical CCD coordinates of the stars change from image to image, so we created the correct parameters that are responsible for referencing the image coordinate system to the WCS (World Coordinate System). We used the console version of the {\scshape astrometry.net} application for this.

Then, using the PHOT task in the {\scshape iraf} package on each frame, we carried out aperture photometry with individual values of the aperture and sky background for each frame. To do this, we used a catalogue of objects generated by means of 2MASS \citep{Skrutskie2006} catalogue and containing equatorial coordinates of stars and their ordinal numbers. Objects were selected basing only on the limiting magnitude $J=14.5$~mag. For the TF1 field the input catalogue contains 23~000 stars, for the TF2 field, 15~000 stars, for the sub-TF2 field, 8500 stars. For the TF3 field the catalogue contains 1000 stars.
 
The radius of the aperture that was used for a particular frame is determined as $0.8\times$ FWHM for the data from the MASTER-II-Ural telescope, and $0.7\times$ FWHM for the RASA telescope. Here FWHM is the mean full width at half maximum of the stars' PSF in the frame. The resulting ratio for an optimal radius of the aperture is obtained empirically from a set of test data for each telescope. 


Next, the obtained data are transferred to the {\scshape astrokit} software. It corrects brightness variations of the stars caused by the variations of atmospheric transparency: for this purpose, for each star in the field of view the software generates an individual ensemble of reference comparison stars which are similar in brightness and position on the frame. Also, {\scshape astrokit} searches for variable stars based on the Robust Median Statistics criterion \citep{Rose2007} which makes it possible to select variable star candidates with more confidence than in the case of analysing the standard deviation of stellar brightness. A more detailed description of the operation of {\scshape astrokit} is provided in \cite{2014AstBu..69..368B}.

The {\scshape k-pipe} script is used separately for the data from each of the target fields in each filter. The output of the script consist of light curves for all the stars of our input catalogue in instrumental stellar magnitudes and variable star candidates.
 
The best photometric precision of the MASTER-II-Ural telescope was attained in the $R_c$ band. For the TF1 field it varied from 0.005 to 0.05~mag for stars with stellar magnitudes from 11 to 16, respectively. For the TF2 field the precision varied from 0.005 to 0.06~mag for stars with stellar magnitudes from 11 to 16, respectively. 

\begin{figure}
\centering
\includegraphics[width=\columnwidth]{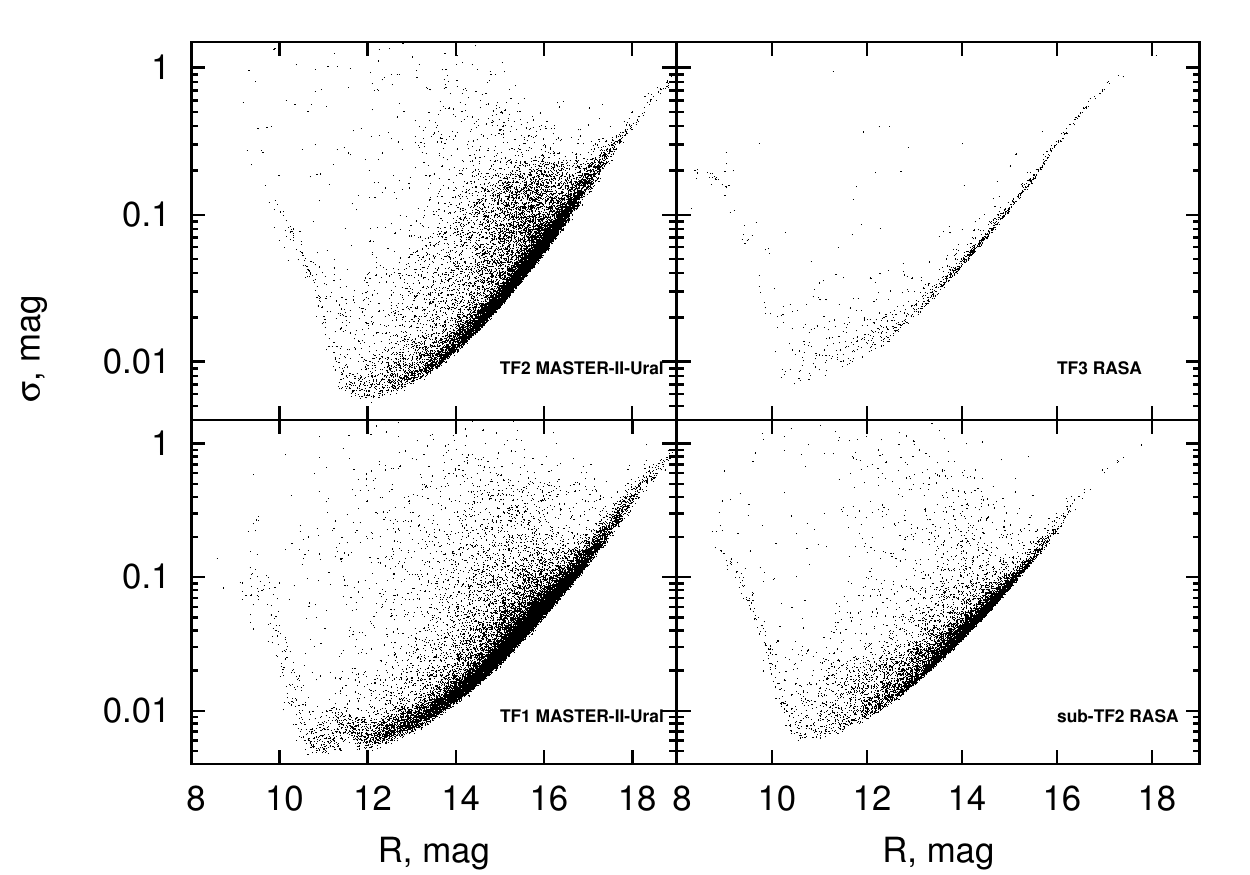}
\caption{Standard deviations of the stars' magnitudes for the entire series of observations vs. their magnitudes in the $R_c$ band.}
\label{figerr}
\end{figure}

The RASA telescope achieved its highest precision in the $R_c$ band for the TF2 field. The precision varied from 0.006 to 0.08~mag for stars with magnitudes ranging between 11 and 16, respectively. The precision of the data for the TF3 field varied from 0.008 to 0.05~mag for stars with stellar magnitudes from 11 to 16, respectively. The precision as a function of magnitude for each field and telescope is presented in Fig.~\ref{figerr}.  

For each dataset we calculated the total number of low-noise stars, i.e. stars whose standard brightness deviation for the entire series of observations was less than 0.02 mag (see table~\ref{tab:low_noise}).

\par
\begin{table}
\small
\caption{Photometric precision of observations in the $R_c$ band obtained with the MASTER-II-Ural and RASA telescopes}
\label{tab:low_noise}
\centering
\begin{tabular}{ccccc}\hline
Field & Telescope & Total number & Low-noise\\
& \phantom{text} & of stars & stars\\
\hline
TF1 & MASTER-II-Ural & 23 500 & 25.5\%\\
TF2 & MASTER-II-Ural & 15 000 & 18.0\%\\
sub-TF2 & RASA & 8 500 & 18.2\%\\ 
TF3 & RASA & 1 000 & 20.9\%\\\hline
\end{tabular}
\end{table}
\par

Due to the highest quantity and accuracy, the data referring to the $R_c$ filter were used to search for transiting hot Jupiter candidates. The frames in the other filters were used to determine the colour indices of the stars.

\subsection{Search for Variable Stars and Exoplanet Transits}

The search for variable stars was carried out by using the algorithm described in \cite{Rose2007}. For each star the Robust Median Statistics (RoMS) criterion $\eta_n$ was calculated: 
\begin{equation}
\begin{gathered}
\eta_n = \frac{\sum\limits_i\displaystyle\frac{\mid m_i - m_{\rm med}\mid}{\sigma_{\rm rms}}}{N-1},
\end{gathered}
\end{equation} where $n$ is the ordinal number of a star; $m_i$ is the $i$-th measurement of the magnitude; $m_{\rm med}$ is the median of
magnitude measurements for star $n$; $N$ is the total number of magnitude measurements for star $n$; $\sigma_{\rm rms}$ is the
estimated standard deviation of star $n$, determined by the least-squares method from the dependence of standard deviation of magnitude on stellar magnitude for the stars in the frame.

The RoMS criterion makes it possible to estimate the magnitude variations of an object. If it exceeds 1, then the star is suspected of variability and is subsequently investigated in greater detail. Based on the results of a visual inspection of the light curves of variable star candidates from the TF1, TF2, and TF3 fields, we were able to find about 500 previously unknown variable stars which are discussed in the following article by \cite{2015PZP....15....7P}. 	

In our search for transiting exoplanets, we selected stars with standard deviations of magnitude for the entire series of observations of less than 0.02~mag (i.e. those for which there is a possibility of detecting a transit of a hot Jupiter with a depth of about 0.01~mag). Prior to searching for transit signals in the light curves, we used the Trend Filtering Algorithm \citep{2005MNRAS.356..557K} in order to remove systematic brightness variations for low-noise stars. The algorithm is used to approximate the light curve for each star in the field by a linear combination of basis vectors (usually, the data for the vectors are provided by the light curves of the stars selected in the field of view). Next, the approximated curve is subtracted from each light curve thus resulting in a photometric time-series partially corrected for common systematic effects. 

The search for specific transit-like signals was carried out by the BLS method (Box-fitting Least Squares) widely used in exoplanet astronomy \citep{2002A&A...391..369K}. This method, as well as the Trend Filtering Algorithm, is implemented in the publicly available {\scshape vartools} Light Curve Analysis Program \citep{2008ApJ...675.1254H}. The detected periods of variability with a high signal-to-noise ratio in the periodogram were used to construct phase-folded light curves in the {\scshape vartools} software which were subsequently analysed visually.

Based on the results derived from analysing the MASTER-II-Ural telescope data, we were able to find two transiting exoplanet candidates with periods less than 1~d in the TF1 field, which subsequently turned out to be astrophysical false positives. The data obtained with the RASA telescope for the TF2 area were analysed separately first, and then together with the data from the MASTER-II-Ural telescope. This way we detected one transiting exoplanet candidate with a period of 1.34~days which is also most likely an astrophysical false positive. In spite of the small number of stars in the TF3 field, we detected a transiting exoplanet candidate with a period of 1.7~d and a transit depth of about 0.01~mag. The nature of this object still remains to be ascertained. 
  
The information about the detected transiting exoplanet candidates is provided in more detail in the next section. 

\section{TRANSITING EXOPLANET CANDIDATES}\label{CANDIDATES}

In our search for transit signals by the BLS method, we found four transiting exoplanet candidates among the low-noise stars; relevant information about the candidates is provided in Table \ref{tab:cands}. 

\par
\begin{table}
\small
\caption{Data on the detected transiting exoplanet candidates}
\label{tab:cands}
\centering
\begin{tabular}{lccccc}\hline
Internal ID & $R_c$ & V & Period & Depth & Duration\\
& mag & mag & d & mag & h\\
\hline
KPS-TF1-3154 & 12.4 & 12.7 & 0.84674 & 0.02 & 1.6\\
KPS-TF1-19251 & 13.9 & 14.3 & 0.98341 & 0.025 & 1.7\\
KPS-TF2-11789 & 14.2 & 14.4 & 1.34652 & 0.02 & 2\\
KPS-TF3-663 & 12.6 & 13.0 & 1.70630 & 0.01 & 1.5\\\hline
\end{tabular}
\end{table}
\par

We carried out additional photometric observations in order to confirm the reality of the visible transits from our initial data. We intended to make sure that the visible decreases in brightness were not caused by systematic errors of observations and/or data processing. By using telescopes with a larger resolving power than that of the wide-field MASTER-II-Ural and RASA telescopes, it was possible to establish whether an exoplanet candidate is a visual binary (multiple) system. The next stage consisted of photometric monitoring of the phase 0.5 to detect possible secondary eclipses, which would reveal the eclipsing binary nature of the candidates. 

The campaign of additional photometric observations involved telescopes with apertures of up to 60~cm, and one 1-m class telescope (T100 of the TUBITAK National Observatory (TUG)). The 6-m BTA telescope of the Special Astrophysical Observatory of RAS was used for spectroscopic and speckle interferometric observations of the candidates. The Kourovka 1.2-m telescope was used for low resolution spectroscopic observations. The information regarding the telescopes is provided in Table \ref{tab:telescopes}. 

The observed data confirmed the detection of actual transits, as well as helped specify the periods of the transits. After that, we carried out spectroscopic and speckle interferometric observations of some of the candidates. The analysis of the data for each of the candidates is reported below.

\begin{table}
\small
\caption{Information about the telescopes that took part in follow-up observations of transiting exoplanet candidates}
\label{tab:telescopes}
\centering
\begin{tabular}{cc}\hline
Telescope & Observatory\\\hline
Celestron EdgeHD & Acton Sky Portal\\
11 SCT & \\
MTM-500 & Kislovodsk Mountain\\
& Astronomical Station\\ 
MASTER-II-Tunka & Irkutsk State University\\
& Astrophysical Centre\\
ZA-320M & Pulkovo Observatory\\
Kreiken & Ankara University\\
& Kreiken Observatory\\
T100 & TUG\\
0.6-m Cassegrain & Torun Centre for \\
& Astronomy Observatory\\
BTA, 6-m & Special Astrophysical\\
& Observatory of RAS\\
KAO-1.2-m & Kourovka Observatory\\
0.43-m Cassegrain & Baronnies Proven\c {c}ales\\
& Observatory\\ 
0.3-m Smith-Cassegrain & Observatori Montcabrer\\
Meade LX200 12'' and 14'' & University of Saskatchewan\\
0.3-m Maksutov-Cassegrain & University of Siena\\
\hline
\end{tabular}
\end{table}

\subsection{KPS-TF1-3154}

\begin{figure}
\centering
\includegraphics[width=\columnwidth]{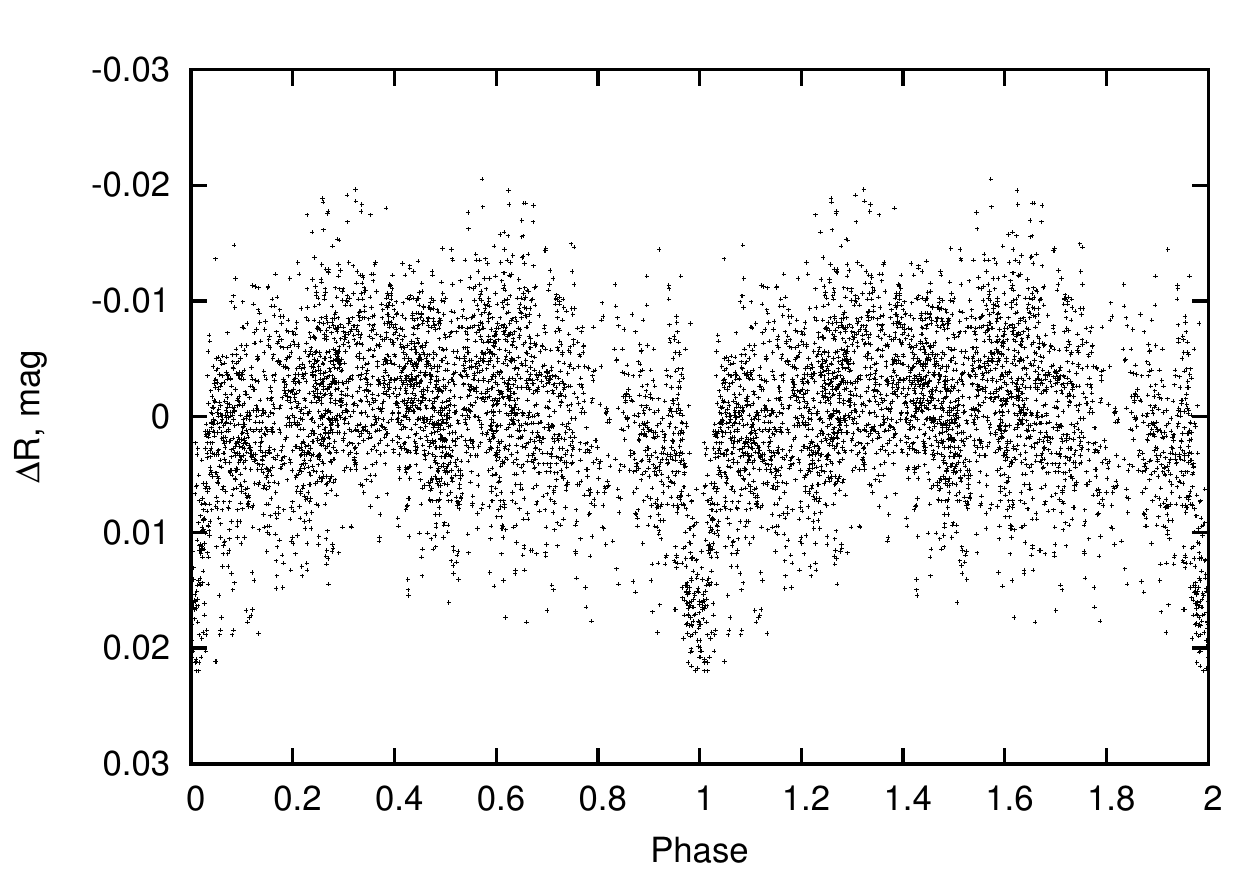}
\caption{Phase-folded light curve of transiting exoplanet candidate KPS-TF1-3154 as obtained using the data from the MASTER-II-Ural telescope.}
\label{fig01}
\end{figure}

The transits of exoplanet candidate KPS-TF1-3154 (labelled as MASTER-1~b in \cite{Burdanov2013}) were observed with the MASTER-II-Ural telescope four times, and three of them were full transits. The BLS method revealed in the periodogram a powerful peak corresponding to a period of 0.84674~d. The light curve folded with the determined period is presented in Fig.~\ref{fig01}. The 2MASS catalogue gives the following magnitudes for the host star 2MASS 20260213+5006032: $J=11.40$~mag, $H=11.05$~mag, and $K_s=10.88$~mag. The initial epoch of the transit minima is $T_{0}(\mathrm{HJD_{TT}})=2456062.826678$.

\begin{figure}
\centering
\includegraphics[width=\columnwidth]{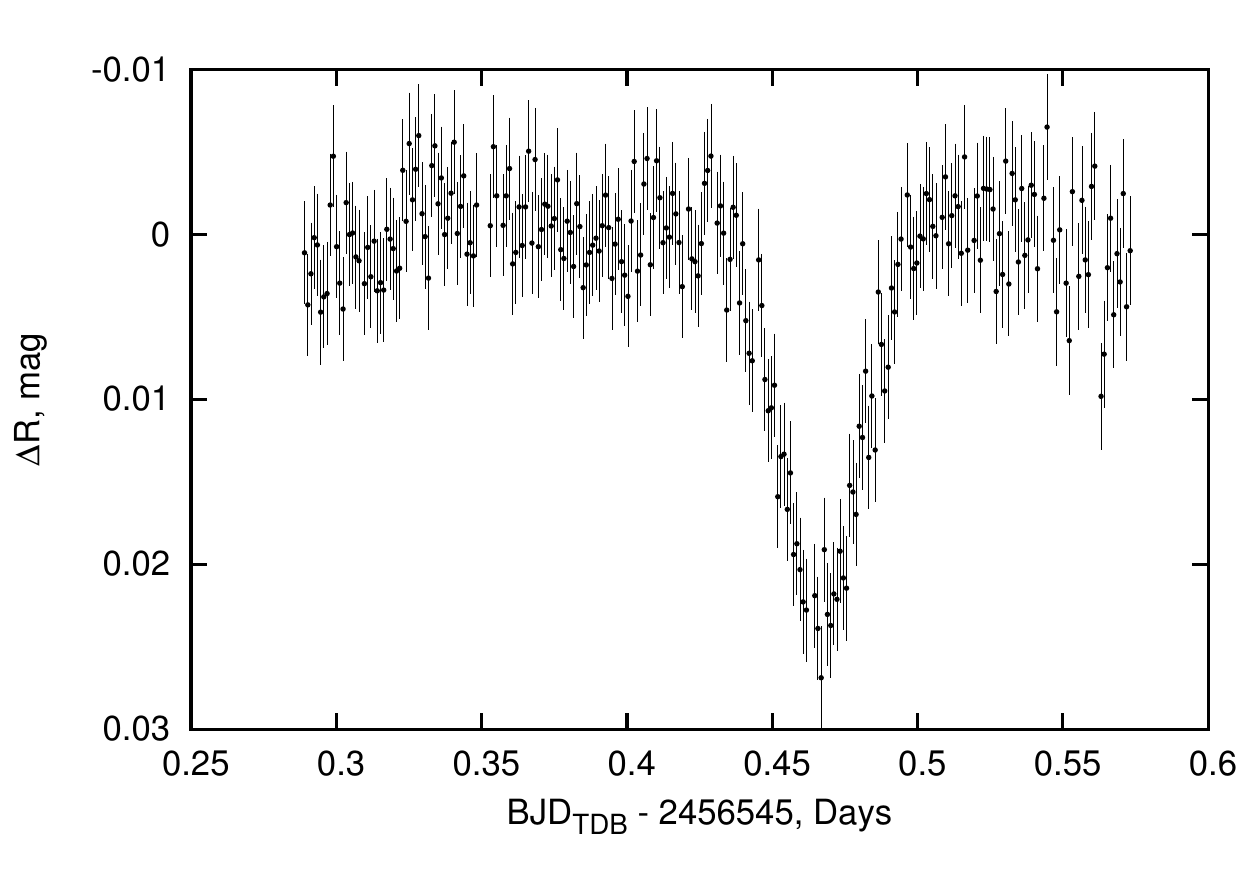}
\caption{Single transit of the exoplanet candidate KPS-TF1-3154 as obtained with the T100 telescope of TUG.}
\label{fig02}
\end{figure}

Additional photometric observations made with the T100 telescope showed a clear V-shape form of the light curve, which is shown in Fig.~\ref{fig02}.

\begin{figure}
\includegraphics[width=\columnwidth]{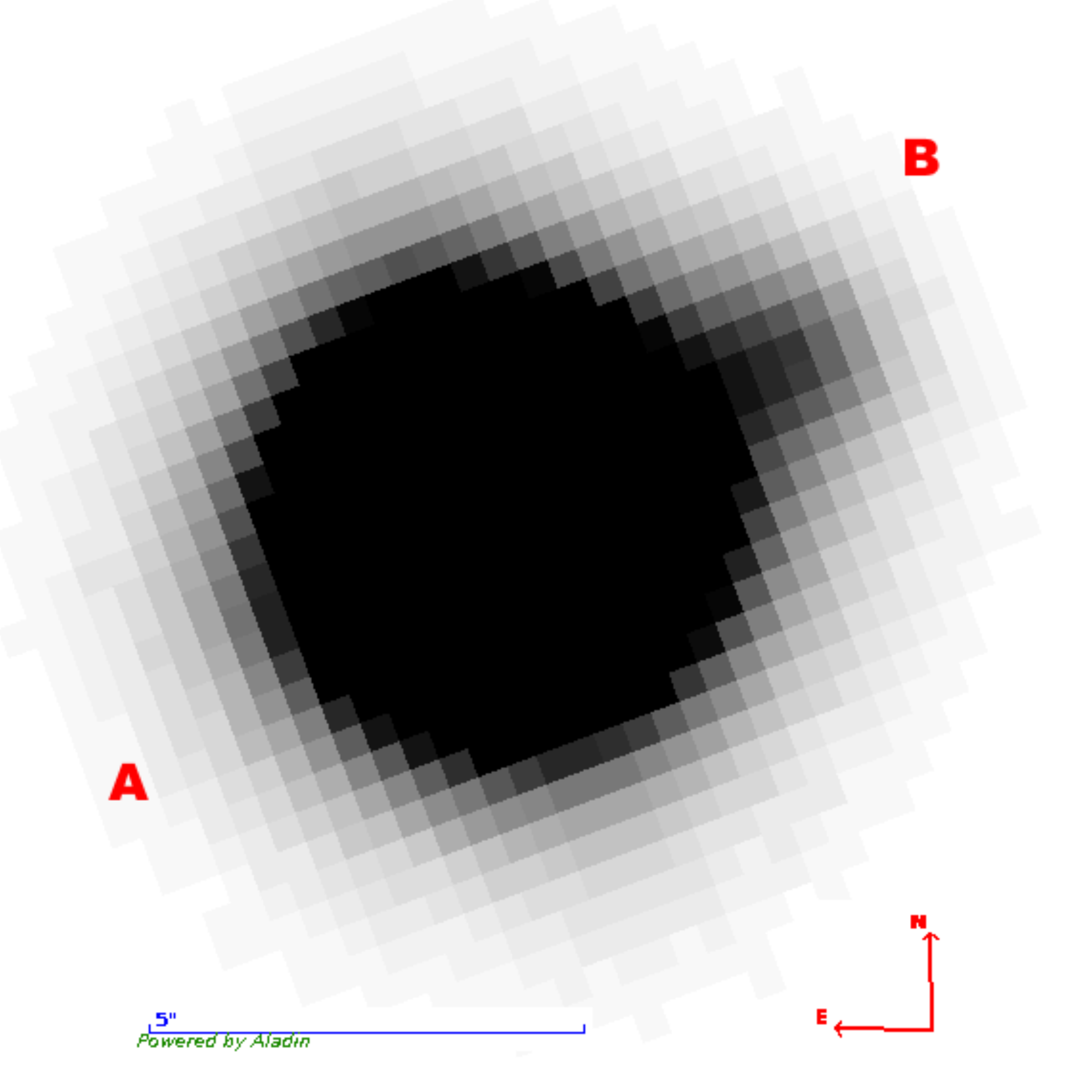}
\caption{Image of transiting exoplanet candidate KPS-TF1-3154 as obtained with the SAO RAS 6-m telescope. The candidate represents a visual binary system.}
\label{fig03}
\end{figure}

On the night of May 17/18, 2013, high angular resolution images and spectra of the KPS-TF1-3154 candidate were obtained with the multi-mode focal reducer SCORPIO-2 \citep{2011BaltA..20..363A} at the prime focus of the SAO RAS 6-m BTA telescope. The FWHM of the image in the SDSS $r$ band is 1.8~arcsec. It was determined that the KPS-TF1-3154 candidate is a visual binary system with the brighter component A and a less bright component B (see Fig.~\ref{fig03}).

To obtain spectra we used the VPHG1200@540 grating and a slit 1~arcsec wide. The image scale along the slit was 0.36~arcsec~pixel$^{-1}$. The spectral resolution of the resulting spectra is about $5~\angstrom$, and the spectral range is $3850-7200~\angstrom$. The slit was oriented along $PA=110\degr$ to obtain spectra from the two components simultaneously. To separate the spectra of a pair of closely related sources we used the following technique. For each column of pixels along the direction across the dispersion, the spectra were approximated by a sum of two Moffat functions and a constant component describing the background. The amplitudes of the functions were represented by the intensity value in the spectrum of the component. The bright component A is a G8V type star as is the fainter component B.

\begin{figure}
\includegraphics[width=\columnwidth]{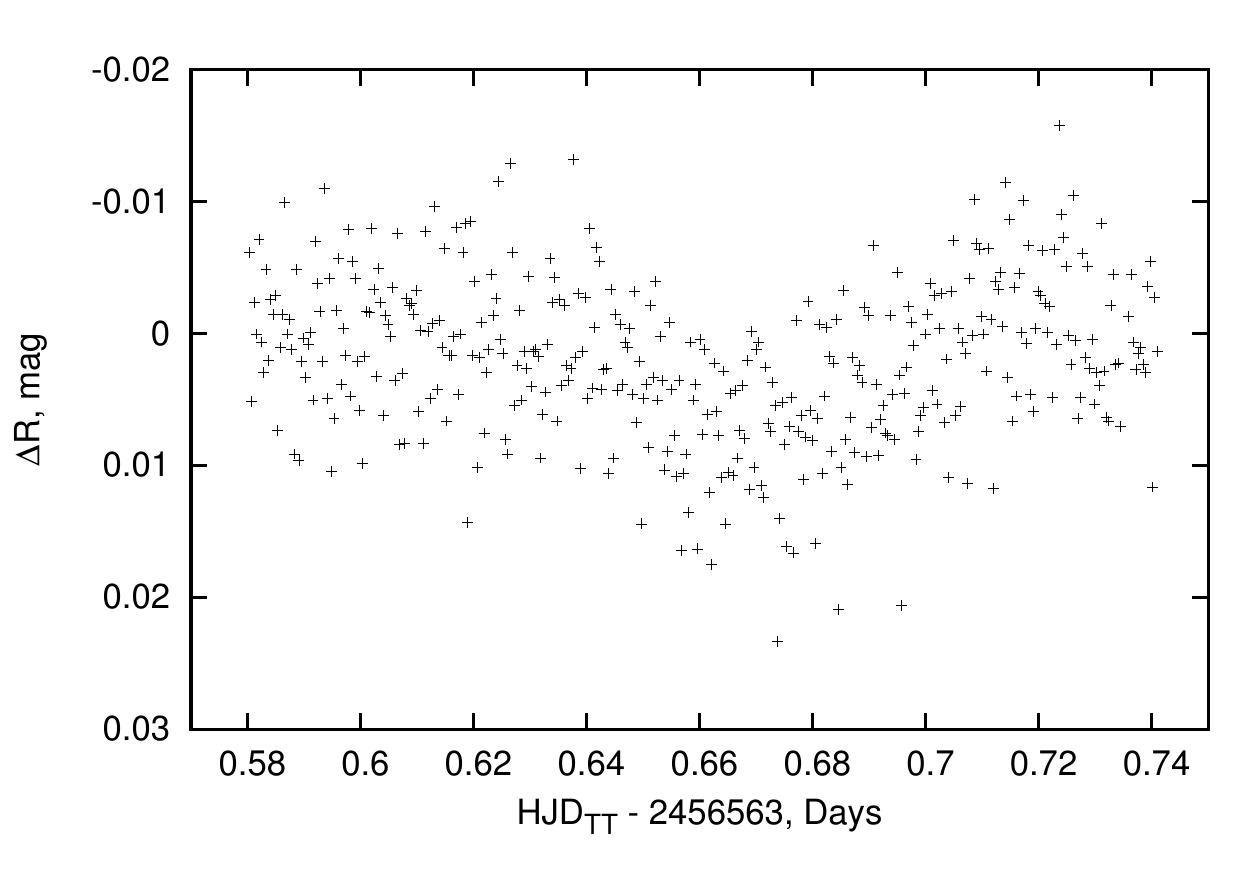}
\caption{Secondary minimum of transiting exoplanet candidate KPS-TF1-3154 as obtained with the EdgeHD 11 SCT telescope (279~mm, f/7) installed at the Acton Sky Portal.}
\label{fig04}
\end{figure}

Subsequent photometric observations at the expected time of secondary minimum revealed a 0.009~mag decrease in brightness lasting for 1.7~h. The observations were made with the EdgeHD 11 SCT telescope (279~mm, f/7) (Celestron Inc., Torrance, CA, USA), installed at the Acton Sky Portal. One of the secondary minima is presented in Fig.~\ref{fig04}. Thus the KPS-TF1-3154 candidate is an astrophysical false positive -- an eclipsing binary system. 

\subsection{KPS-TF1-19251}

\begin{figure}
\includegraphics[width=\columnwidth]{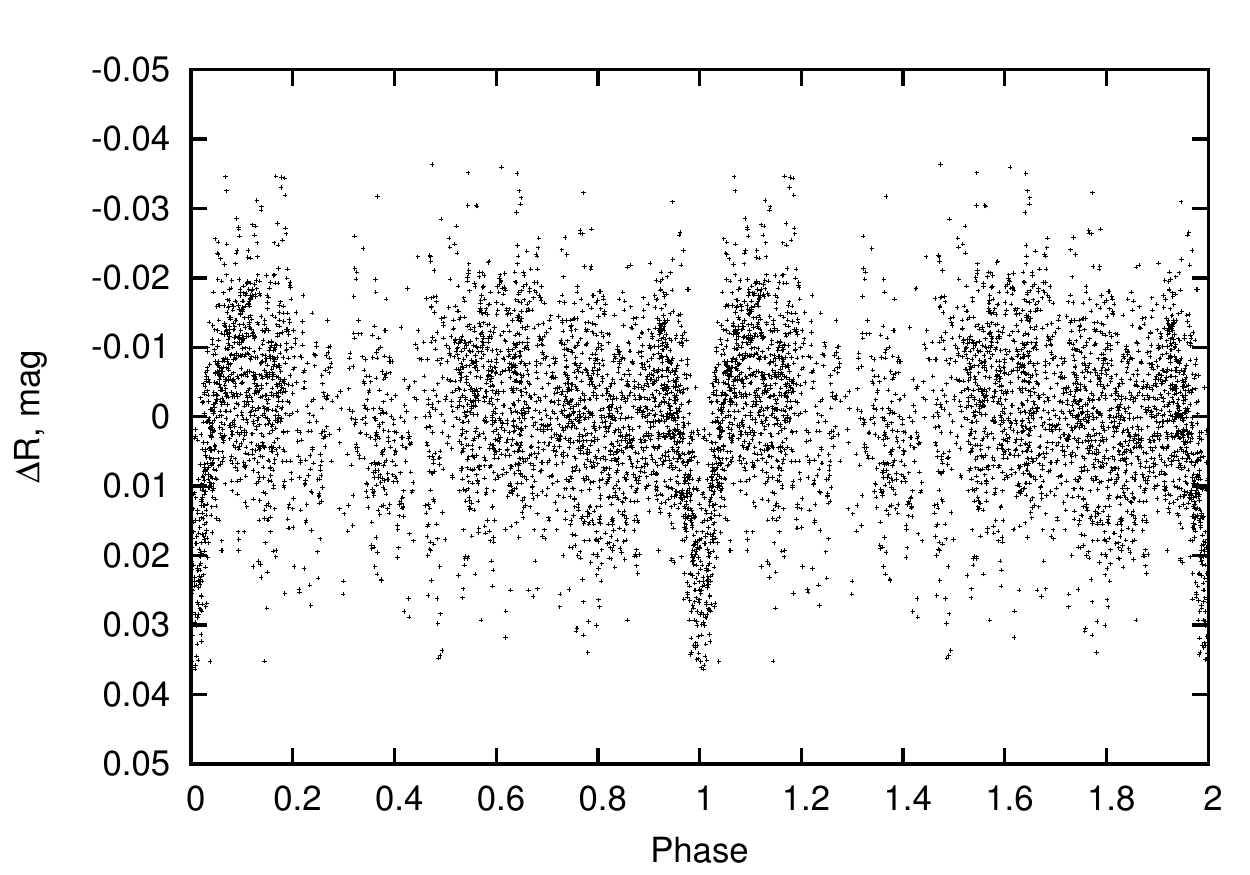}
\caption{Phase folded light curve of transiting exoplanet candidate KPS-TF1-19251 as obtained with the MASTER-II-Ural telescope.}
\label{fig05}
\end{figure}

Four transits of the exoplanet candidate KPS-TF1-19251 were observed with the MASTER-II-Ural telescope; two of them were full transits. The BLS method revealed in the periodogram a powerful peak corresponding to a period of 0.98341~d. The resulting light curve folded with the determined period is presented in Fig~\ref{fig05}. The magnitudes of the host star 2MASS 20341625+5015427 are $J=12.60$~mag, $H=12.21$~mag, and $K_s=12.10$~mag. The initial epoch of the transit minima is $T_{0}(\mathrm{HJD_{TT}})=2456062.956109$.

\begin{figure}
\includegraphics[width=\columnwidth]{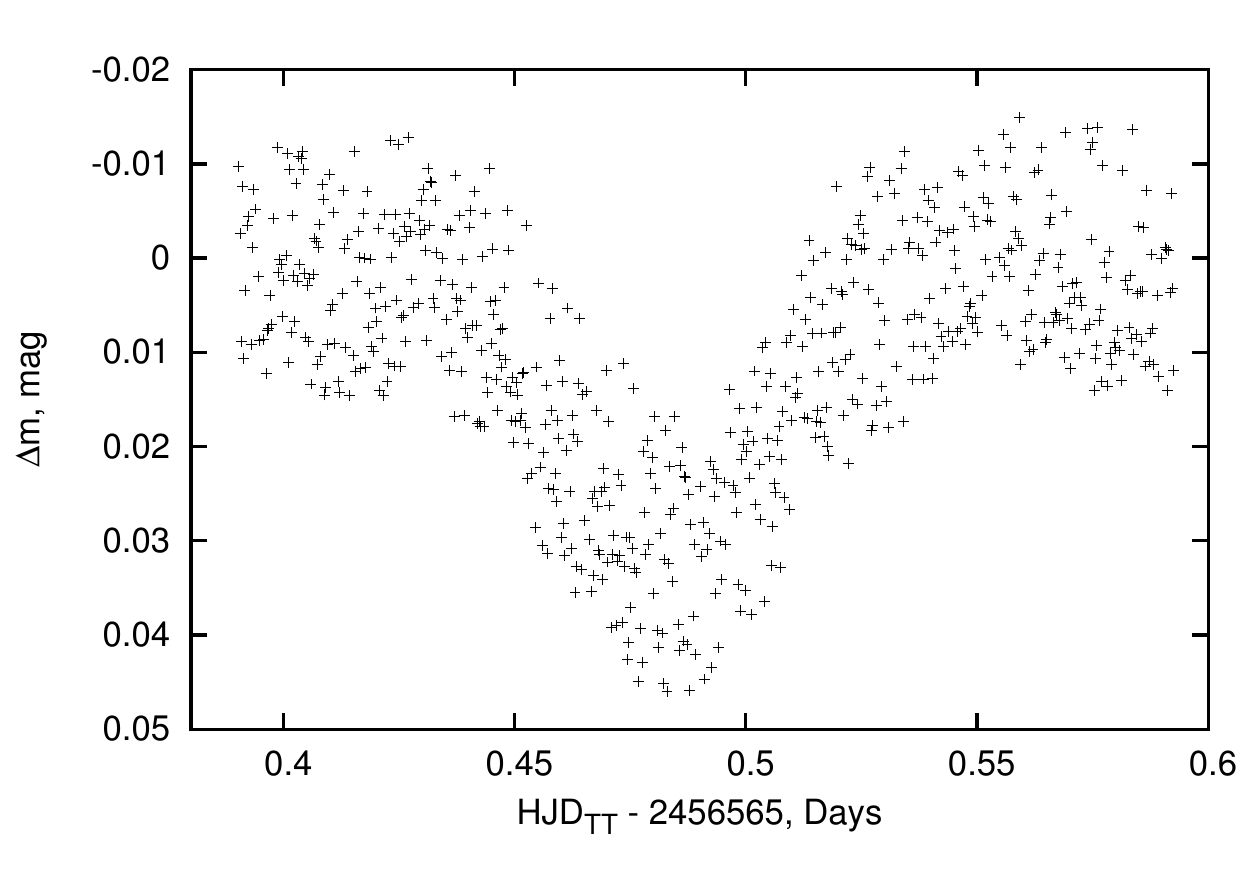}
\caption{Single transit of the exoplanet candidate KPS-TF1-19251 as obtained with the 0.6-m Cassegrain telescope at the Centre for Astronomy of the Nicolaus Copernicus University in Torun (Poland).}
\label{fig06}
\end{figure}

Additional photometric observations revealed a V-shape form of the light curve with the transit depth 0.025~mag and duration 2.5~h. One such transit, obtained with the 60-cm telescope of the Centre for Astronomy of the Nicolaus Copernicus University, is presented in Fig.~\ref{fig06}.

Decreases in brightness at the times corresponding to secondary eclipses were not revealed (from photometric observations with 0.003~mag precision).

\begin{figure}
\includegraphics[width=\columnwidth]{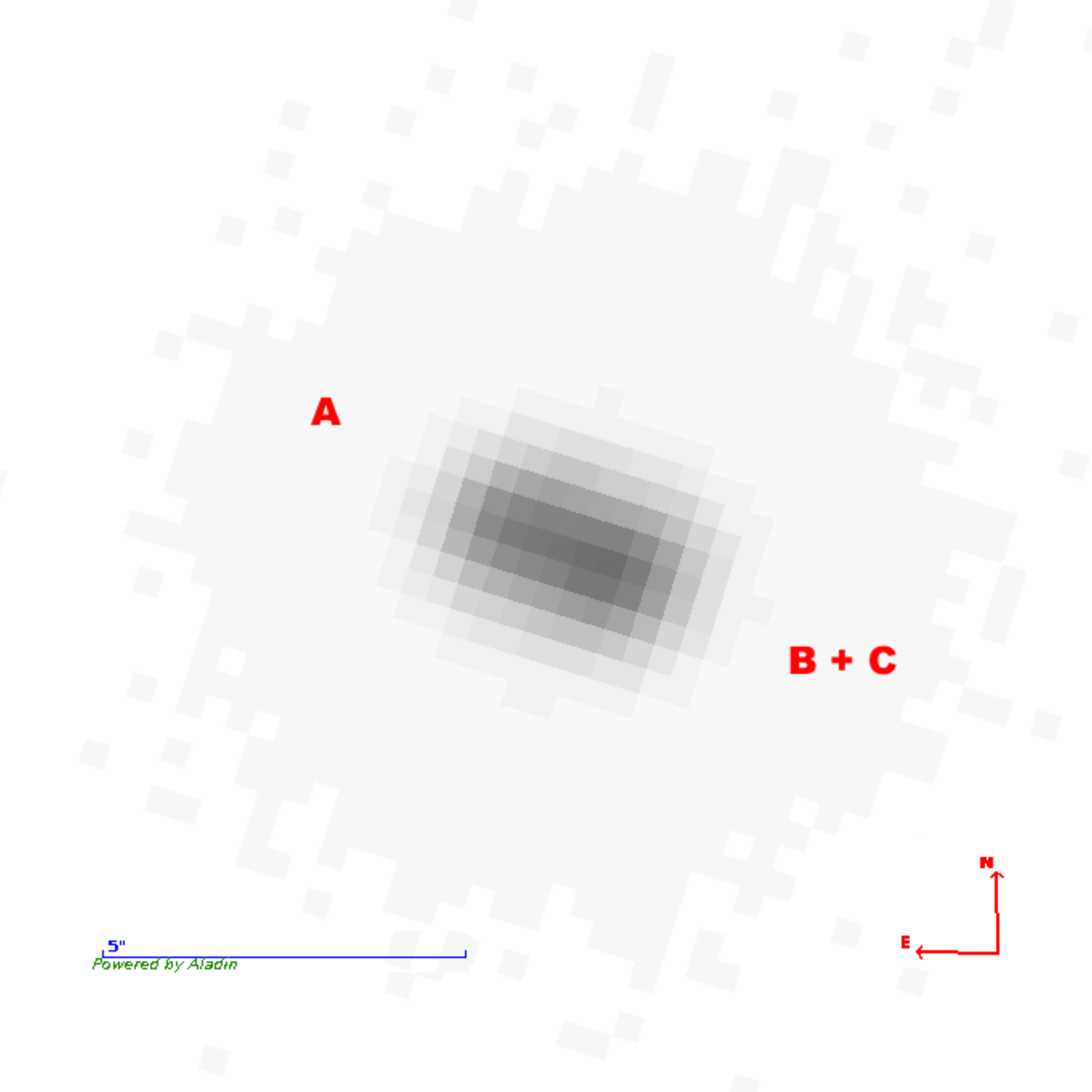}
\caption{High angular resolution image of transiting exoplanet candidate KPS-TF1-19251 as obtained with the SAO RAS 6-meter telescope. According to these data, the candidate represents a visual binary system. Actually, it is a triple system (see Fig.~\ref{fig08}).}
\label{fig07}
\end{figure}

On the same observation night as for the KPS-TF1-3154 candidate, we obtained high angular resolution images and spectra of KPS-TF1-19251 with SCORPIO-2 at the SAO RAS 6-m BTA telescope. The seeing was about 1.2~arcsec. We found that the KPS-TF1-19251 candidate is also a visual binary system consisting of two components of almost identical brightness and separated by 1.3~arcsec (see Fig.~\ref{fig07}). Based on spectral data, we found that the brighter component A is a dwarf of class G2 and the weaker component B is a dwarf of class K0. 

\begin{figure}
\includegraphics[width=\columnwidth]{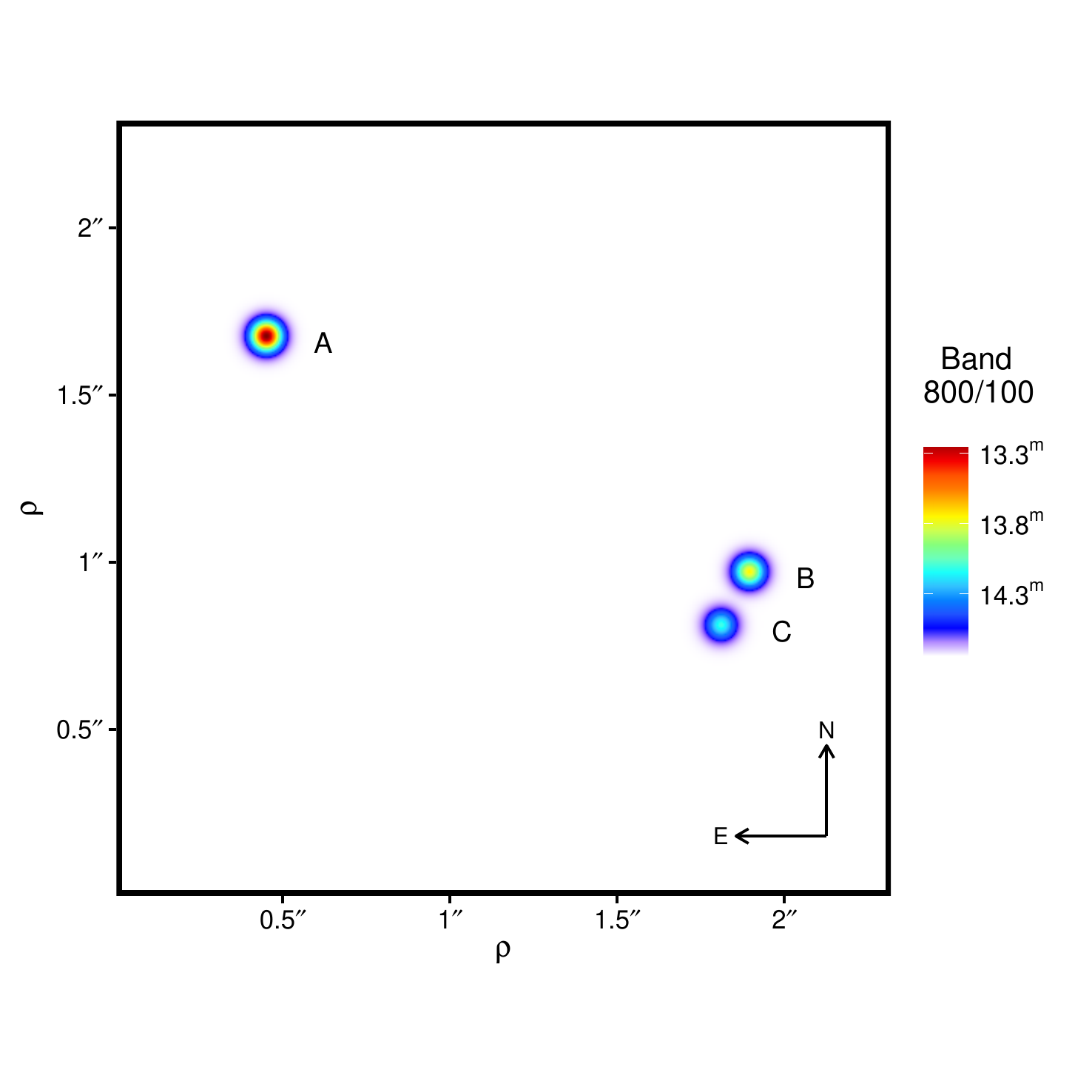}
\caption{Model of transiting exoplanet candidate KPS-TF1-19251 based on speckle interferometric images obtained with the SAO RAS 6-meter BTA telescope.}
\label{fig08}
\end{figure}

On September 05, 2014, the SAO RAS 6-m BTA telescope carried out speckle interferometric observations of transiting exoplanet candidate KPS-TF1-19251. The speckle interferometer acquired six series of short-exposure images with x~10 and x~16 micro-objective lenses in the 600/40, 800/100, and 900/80~nm bands \citep{2009AstBu..64..296M}. The entire system was observed in a single field of view. It was discovered that the system's faint component B is a speckle binary system and contains another component, C. The position parameters of the system were determined from the image series obtained in the 800/100~nm band. The angular distances between the system's components are: $\rho_{AB}=1317~\pm~14$~mas and $\rho_{BC}=148~\pm~12$~mas; the position angles are $\Theta_{AB}=244\degr~\pm~1\fdeg5$ and $\Theta_{BC}=152\degr~\pm~2\degr$. Because of the low signal-to-noise ratio, we were unable to determine the parameters from the series in the 600/40 and 900/80~nm filters. The brightness differences in the 800/100~nm band are: $\Delta m_{(A - (B+C))}=0.01~\pm~0.05$~mag and $\Delta m_{(B - C)}=0.5~\pm~0.5$~mag. The obtained image is presented in Fig.~\ref{fig08}. Based on these data, we estimate the spectral class of component C as K5.

The available data are insufficient to determine which component of this triple system has periodic transits. But sequential estimates were made of the 'true' depths of the eclipses for each of the components of this visual triple system. The 'true' depth of an eclipse is defined as the decrease in brightness of one of the components which would be observable in the absence of other stars in the aperture. If eclipses occur for star A, then the true depth of an eclipse would be about 0.05~mag. Such decreases in brightness are not typical for transits of extrasolar planets, even for the most inflated exoplanet orbiting a main sequence star known to date (WASP-17 b, see \cite{2010ApJ...709..159A}). If we assume that a 0.05~mag decrease in brightness is caused by a planet orbiting a G2V star, then its radius would be equal to 2.2~radii of Jupiter. If eclipses occur for star B or C, then the true depths of the eclipses would be equal to 0.08 or 0.14~mag, respectively. Such a depth is also too large for the eclipsing object to be an exoplanet. Taking into account the V-shaped form of the light curve, the system is most likely an eclipsing binary star. Although the discrepancy between the hot Jupiters' predicted radii and observed radii (known as 'radius anomaly') has not yet been solved, there still might be a chance of an exoplanet companion orbiting star~A. Of course, the nature of the eclipsing body can be ascertained conclusively by using radial velocity observations, i.e. by estimating its mass. 

\subsection{KPS-TF2-11789}

\begin{figure}
\includegraphics[width=\columnwidth]{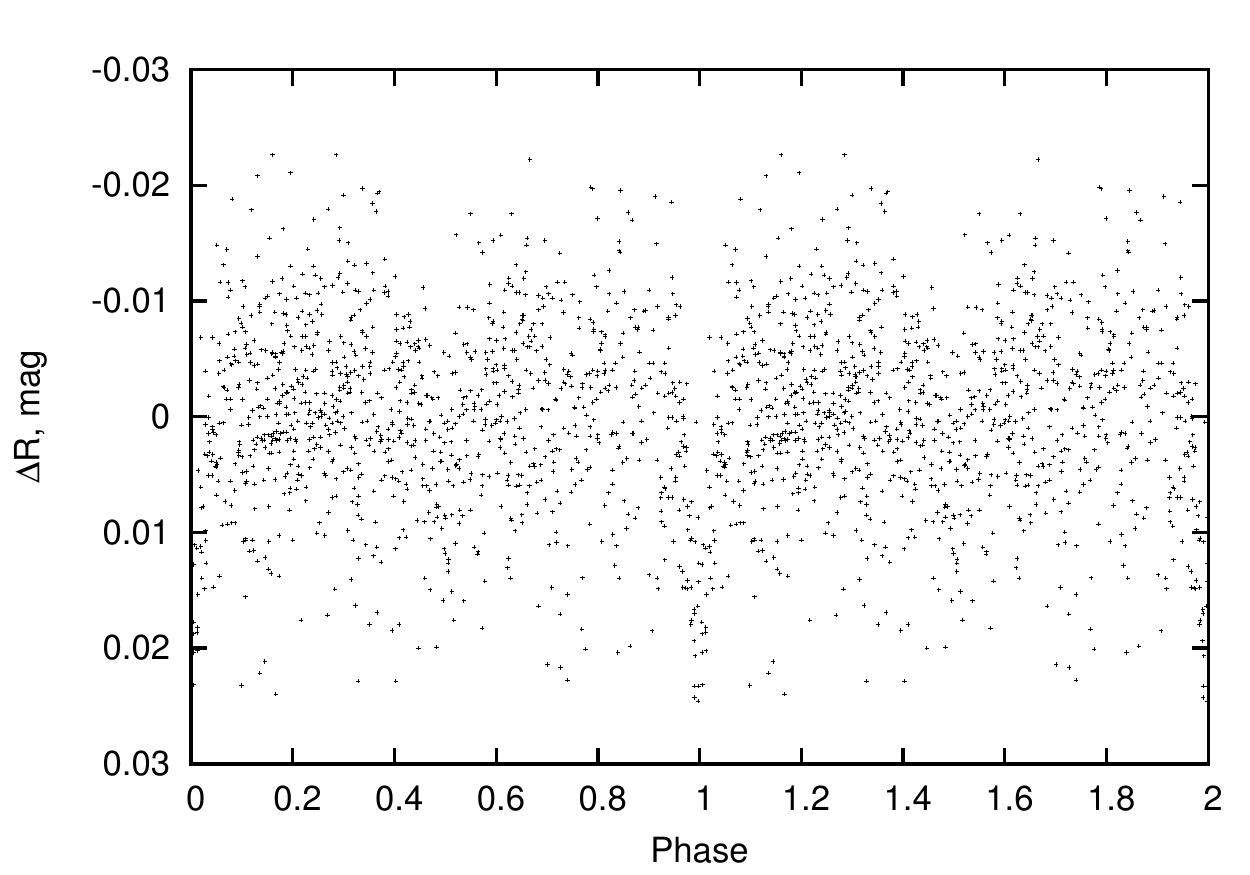}
\caption{Phase folded light curve of transiting exoplanet candidate KPS-TF2-11789 as obtained from the combined data of the MASTER-II-Ural and RASA telescopes. The main minimum is clearly evident, and a hint of a secondary minimum is also visible.} 
\label{fig09}
\end{figure}

Transiting exoplanet candidate KPS-TF2-11789 was discovered in the TF2 field of the KPS survey. The transits were observed with the MASTER-II-Ural telescope and the RASA telescope. In the combined dataset, the BLS method revealed a powerful peak in the periodogram, but now corresponding to a longer period of 1.34652~d. The resulting phase curve folded with the period thus determined is presented in Fig.~\ref{fig09}. According to the 2MASS catalogue, the magnitudes of the host star 2MASS 02472435+6324230 are $J=13.04$~mag, $H=12.84$~mag, and $K_s=12.74$~mag. The initial epoch of the transit minima is $T_{0}(\mathrm{HJD_{TT}})=2456365.223054$.

\begin{figure}
\includegraphics[width=\columnwidth]{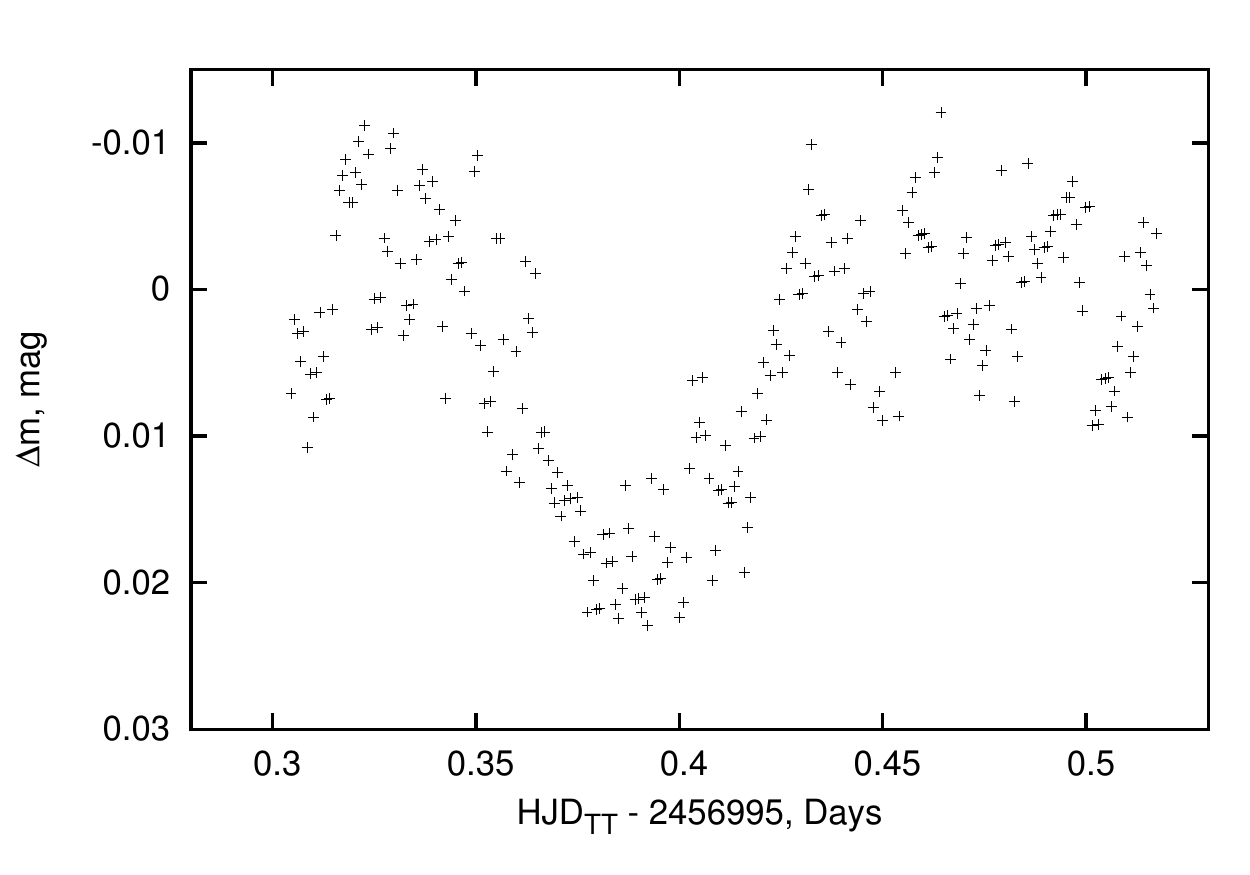}
\caption{Single transit of the KPS-TF2-11789 exoplanet candidate as obtained with the MTM-500 telescope.} 
\label{fig10}
\end{figure}

\begin{figure}
\includegraphics[width=\columnwidth]{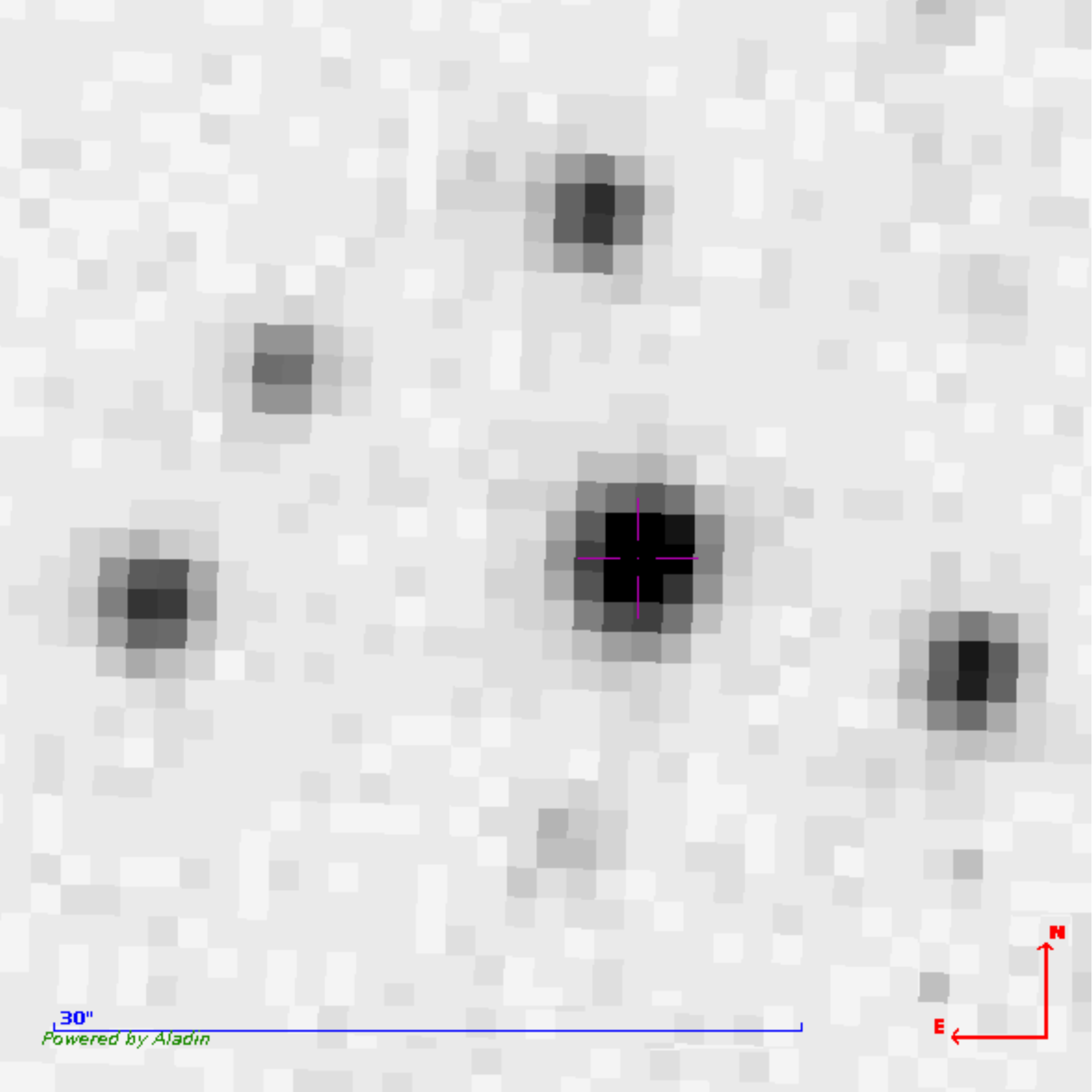}
\caption{Image of transiting exoplanet candidate KPS-TF2-11789 as obtained with the MTM-500 telescope.} 
\label{fig11}
\end{figure}

As in the case with the previous candidates, additional photometric observations were made in order to specify the shape of the light curve, as well as to check for the presence of a secondary minimum. The MTM-500 telescope (500~mm, f/8.2) of the Kislovodsk Mountain Astronomical Station was used to obtain several transit light curves (one of them is plotted in Fig.~\ref{fig10}). It is a U-shaped curve, the transit depth is 0.02~mag, and the duration of the transit is about 1.8~h. It was not possible to detect any visual companions of the KPS-TF2-11789 candidate from the images available (see Fig.~\ref{fig11}).

\begin{figure}
\includegraphics[width=\columnwidth]{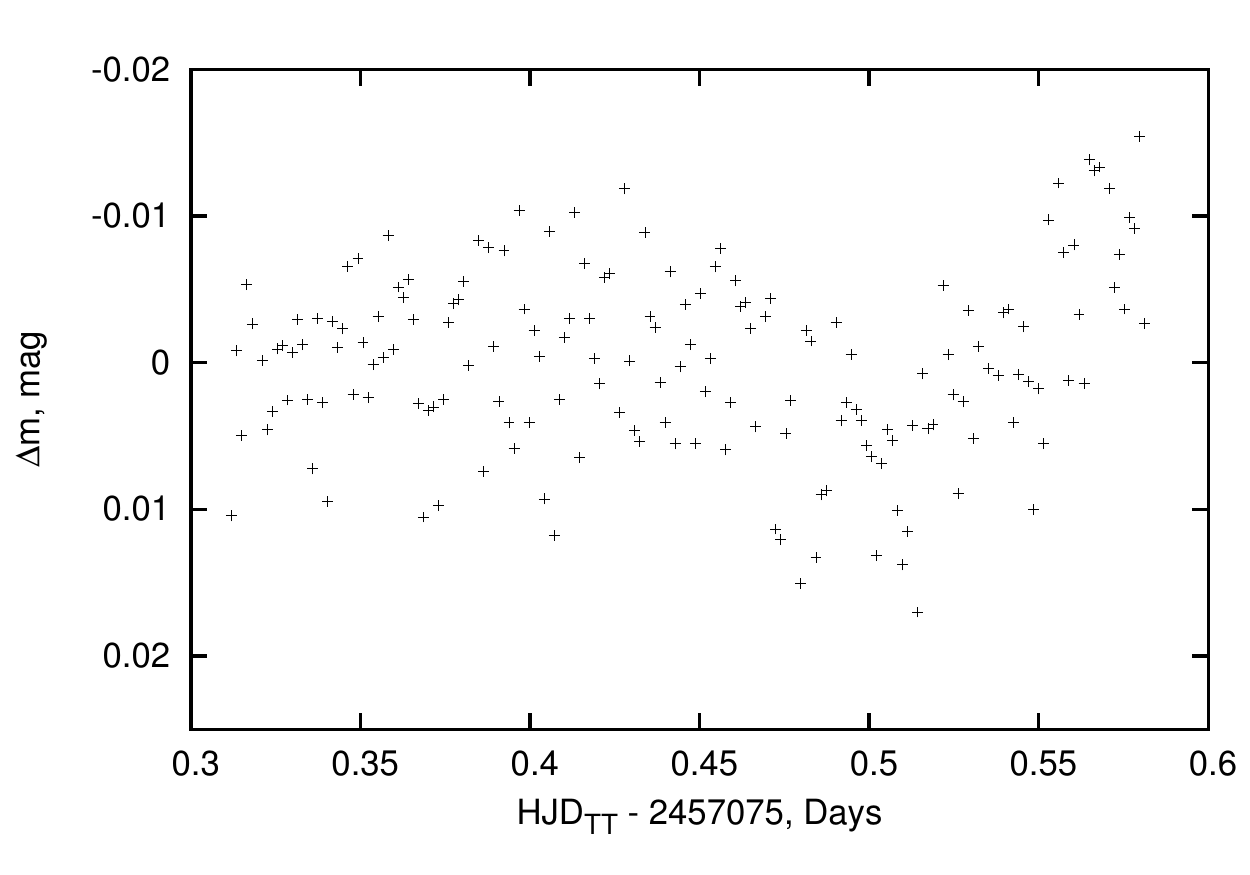}
\caption{Secondary minimum of transiting exoplanet candidate KPS-TF2-11789 as obtained with the MTM-500 telescope. The theoretical time of the minimum (JD~2457075.51293) coincides with the minimum on the light curve.} 
\label{fig12}
\end{figure}

However, it was possible to detect a secondary minimum with a depth of about 0.006~mag and a duration of about 2~h at the predicted time. One such light curve with a secondary minimum, obtained with the MTM-500 telescope, is plotted in Fig.~\ref{fig12}. Thus, in all likelihood, this candidate is also an astrophysical false positive, namely an eclipsing variable star. 

The spectral class of the system was determined as F0 through the use of low resolution spectra obtained with the ANNA spectrograph of the 1.2-m telescope at the Kourovka astronomical observatory.

\subsection{KPS-TF3-663}

\begin{figure}
\includegraphics[width=\columnwidth]{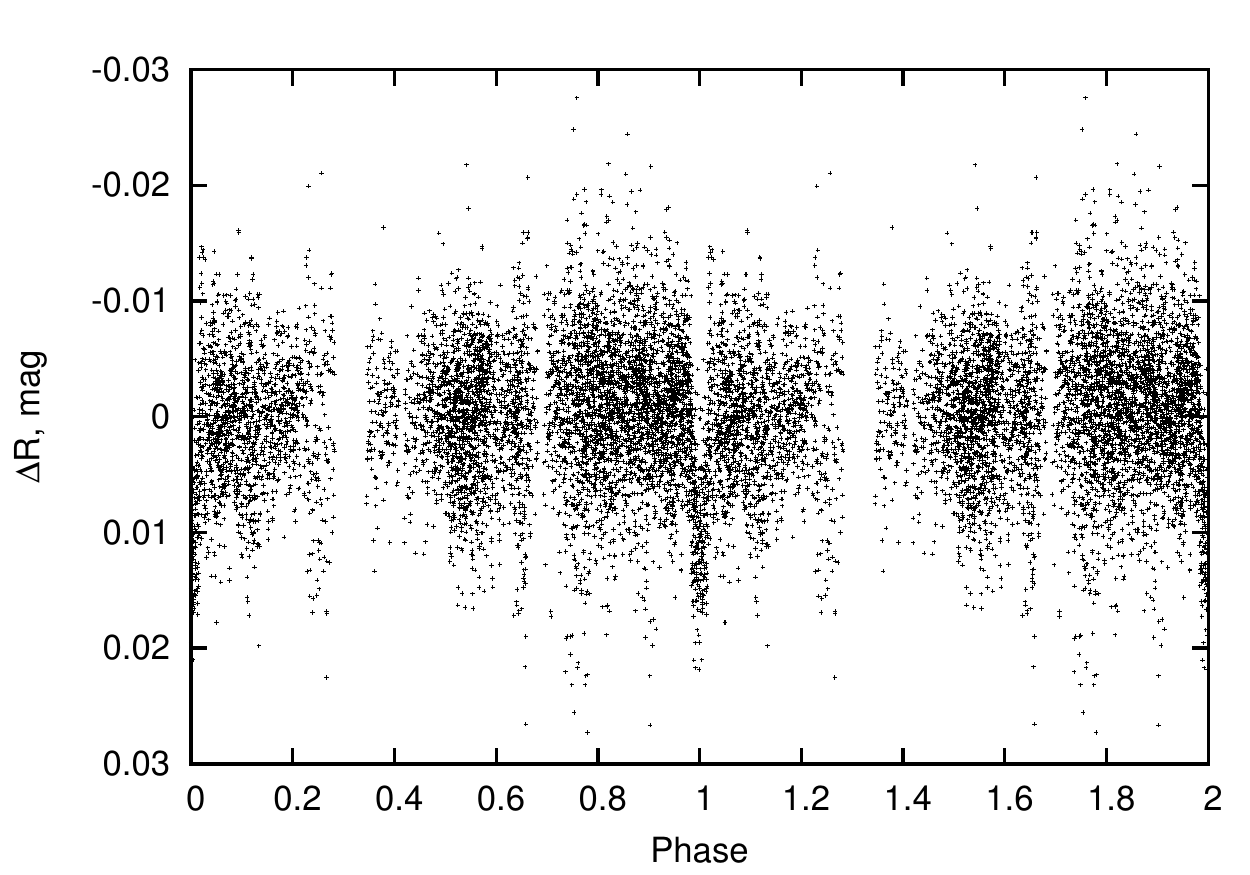}
\caption{Phase folded light curve of transiting exoplanet candidate KPS-TF3-663 as obtained from the RASA telescope data.} 
\label{fig13}
\end{figure}

Initially, the transits of the exoplanet candidate KPS-TF3-663 were observed only with the RASA telescope. We detected five transits, and two of them were full transits. The BLS method revealed in the periodogram a peak corresponding to a period of 1.70630~d. The resulting light curve folded with the determined period is presented in Fig~\ref{fig13}.

\begin{figure}
\includegraphics[width=\columnwidth]{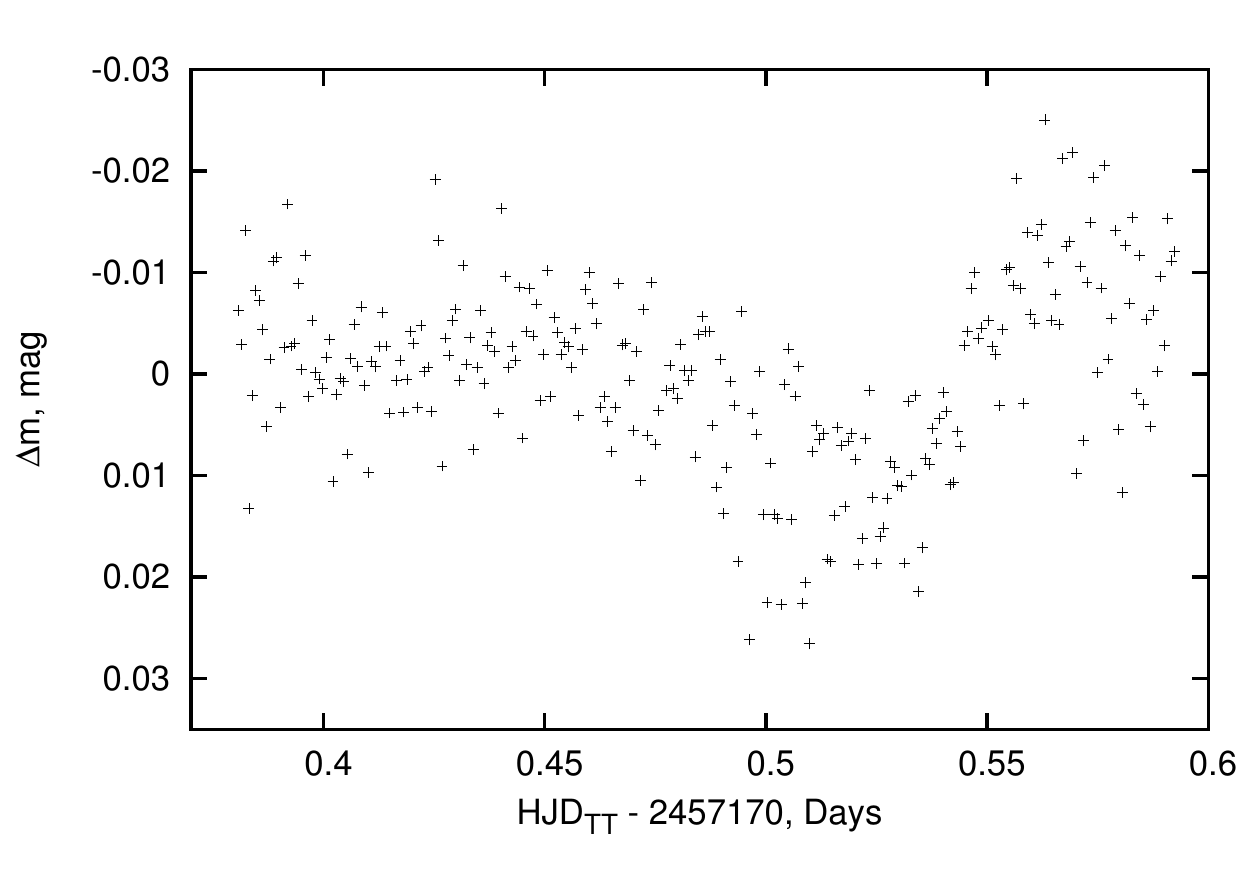}
\caption{Single transit of the exoplanet candidate KPS-TF3-663 as obtained by the Baronnies Proven\c {c}ales Observatory telescope.}
\label{fig14}
\end{figure}

One of the best transits of the KPS-TF3-663 candidate in terms of precision, shown in Fig.~\ref{fig14}, was acquired by the 0.43-m Cassegrain telescope located at the Baronnies Proven\c {c}ales Observatory. It is a U-shaped curve, the transit depth is 0.01~mag, and the duration of the transit is about 1.5~h. Decreases in brightness at the predicted times corresponding to a secondary minimum were not detected (from photometric observations with 0.003~mag precision). 

On June 08, 2015, the SAO RAS 6-m BTA telescope carried out speckle interferometric observations of the KPS-TF3-663 candidate. We were unable to detect a visual companion of the candidate from these data.

According to the MASTER-II-Ural telescope observations, the magnitudes of the candidate are: $B=14.1$, $V=13.0$, $R=12.6$, and $I=12.8$~mag. Taking into account JHK photometry from the 2MASS catalogue ($J=11.4$, $H=11.0$, and $K_s=10.9$~mag), we suggest that the spectral type of the star is G8--K1.

KPS-TF3-663 is the most promising exoplanet candidate, and spectral observations are needed to clarify the nature of the eclipsing body.

\section{CONCLUSIONS}\label{CONCL}

We have presented the first results of the Kourovka Planet Search project. Over a period of two years, the MASTER-II-Ural telescope observed two fields of size $2\times2$~deg$^2$ (TF1 and TF2), and the RASA telescope made observations of the TF2 subregion of size $1.2\times1.6$~deg$^2$ as well as of the TF3 field. 

As a result, four transiting exoplanet candidates and about 500 new variable stars were discovered among the 39~000 stars of the input catalogue. The transiting exoplanet candidates thus discovered were all identified reliably by the BLS method. Three of the four candidates represent one of the most widespread types of astrophysical false positives. The nature of the fourth candidate, which is the most promising one, can be ascertained by further spectral observations. We would like to conclude that amateur wide field observations, such as the RASA setup at Acton Sky Portal, can provide high quality data for discovering exoplanet candidates. 
  
We associate the lack of discovered transiting hot Jupiters with a low duty cycle of observations made by a single instrument and with the small number of observed fields. Unfortunately, poor astroclimatic conditions at the Kourovka astronomical observatory result in only $\approx~3$~h (on average) of observations a night.  Also, in the course of our observations, the TF2 field was first observed with the MASTER-II-Ural telescope, and then with the RASA telescope only. The estimates of our exoplanet survey yield (with the real duty cycle and number of low-noise stars) the detection of less than 0.5 hot Jupiters. In the future, we would like to relocate the MASTER-II-Ural telescope to a site with a better astroclimate, and observe identical fields of the sky with the two telescopes separated in longitude. 

We assume that the relatively small number of discovered transiting exoplanet candidates, compared to the wider-field surveys, might be due to the fact that our observations encompassed a smaller number of fields in the sky and that the image scale was different (1.8~arcsec~pixel$^{-1}$ vs. 14~arcsec~pixel$^{-1}$ in the case of the WASP and HATNet surveys). This smaller image scale results in a smaller number of blending stars and, as a consequence, a smaller number of astrophysical false positives. 
 
Another way of improving the probability of detecting transiting exoplanets involves observing a larger number of fields in the sky. Both telescopes were used to make trial observations in an alternative (cyclic) mode: during a single night, observations of three to five adjacent image fields of the celestial sphere were made, one of which included a known transiting exoplanet. The transit light curve obtained in such a mode of observation has a lower time resolution, but the transit phase is clearly evident and the BLS method allows confident discovery of the transit signal. In future observation campaigns, we plan to observe three to five fields with the two telescopes every night.

\section*{Acknowledgements}
This work was partly supported by the Russian Foundation for Basic Research project No.~14-02-31338, by Act 211 of the Government of the Russian Federation (contract No.~02.A03.21.0006), and by the Ministry of Education and Science of the Russian Federation (the basic part of the state assignment, registration number 01201465056). Some of the observations in this work were obtained using the equipment of the unique scientific facility 'Kourovka astronomical observatory', and was supported by the Russian government through the Ministry of Education and Science of the Russian Federation (unique grant identifier RFMEFI59114X0003). Evgenii N. Sokov acknowledges the support of the Russian Science Foundation grant No.~14-50-00043 in conducting an international photometric observing campaign of the discovered transiting exoplanet candidates, the Russian Foundation for Basic Research (project No.~14-02-92615~KO~a), the UK Royal Society International Exchange grant~IE140055, and the programmes of the Presidium of the Russian Academy of Sciences~(P9). KI acknowledges the support by state order No.~3.615.2014/K of the Russian Ministry of Science and Education. Observations with the 6-m telescope of SAO RAS were carried out with the financial support of the Ministry of Education and Science of the Russian Federation (agreement No.~14.619.21.0004, project ID~RFMEFI61914X0004). AM is also grateful for the financial support of the 'Dynasty' Foundation. The speckle interferometric observations of the exoplanet candidates with the 6-m telescope of SAO RAS were supported by the Russian Science Foundation grant No.~14-50-00043. \"OB and I\"O thank TUBITAK for partial support in using the T100 telescope (project No.~12CT100-378). 

The authors would like to thank Dr.~M.~Gillon and S.~T.~F.~Padovani for their valuable comments, and I.~Sergey for his attempts to conduct photometric follow-up observations of the discovered candidates. 

\bibliography{Burdanov}


%

\end{document}

%% file: journal.tex
\def\apj{Astrophys.~J}
\def\apjl{Astrophys.~J.,~Lett}
\def\apjs{Astrophys.~J.,~Suppl.~Ser}

\def\actaa{Acta Astron}
\def\aap{Astron.~Astrophys}

\def\aj{Astron.~J}

\def\mnras{Mon.~Not.~R.~Astron.~Soc}

\def\pasp{Publ.~Astron.~Soc.~Pac}

